\documentclass[amsmath,amssymb,superscriptaddress,nobalancelastpage,prb,twocolumn]{revtex4-1}
\usepackage{blindtext} 
\usepackage{graphicx}
\usepackage{color}

\newcommand{\Sr}{Sr$_{2}$RuO$_4$}

\newcommand{\Ca}{Ca$_{2}$RuO$_4$}
\newcommand{\ketxy}{$d_{xy}^\pm$}%{$\rvert xy, \pm\rangle$}
\newcommand{\ketyz}{$d_{yz}^\pm$}%{$\rvert yz, \pm\rangle$}
\newcommand{\ketxz}{$d_{xz}^\pm$}%{$\rvert xz, \pm\rangle$}
\newcommand{\ketyzmp}{$d_{yz}^\mp$}%{$\rvert yz, \pm\rangle$}
\newcommand{\ketxzmp}{$d_{xz}^\mp$}%{$\rvert xz, \pm\rangle$}
\newcommand{\ketone}{$\psi_{1}^\pm$}%{$\rvert 0, \pm\rangle$}
\newcommand{\kettwo}{$\psi_{2}^\pm$}%{$\rvert 1, \pm\rangle$}
\newcommand{\ketthree}{$\psi_{3}^\pm$}%{$\rvert 2, \pm\rangle$}
\begin{document}

%opening

%\author{}
%\title{Spin-orbit interaction and crystal field splitting effects on the electronic structure of {(Ca,Sr)}$_2$RuO$_4$} %: A resonant inelastic x-ray study}
%\title{Spin-orbit coupling and orbital excitations in {Sr}$_2$RuO$_4$ and {Ca}$_2$RuO$_4$}
\title{Spin-Orbit-Induced Orbital Excitations in {Sr}$_2$RuO$_4$ and {Ca}$_2$RuO$_4$: A Resonant Inelastic X-ray Scattering Study }

\author{C. G. Fatuzzo}
\affiliation{Institute for Condensed Matter Physics, \'{E}cole Polytechnique Fed\'{e}rale
de Lausanne (EPFL), CH-1015 Lausanne, Switzerland}

\author{M. Dantz}
\affiliation{Swiss Light Source, Paul Scherrer Institut, CH-5232 Villigen PSI, Switzerland}

\author{S. Fatale}
\affiliation{Institute for Condensed Matter Physics, \'{E}cole Polytechnique Fed\'{e}rale
de Lausanne (EPFL), CH-1015 Lausanne, Switzerland}

\author{P.~Olalde-Velasco}
\affiliation{Swiss Light Source, Paul Scherrer Institut, CH-5232 Villigen PSI, Switzerland}

\author{N.~E.~Shaik}
\affiliation{Institute for Condensed Matter Physics, \'{E}cole Polytechnique Fed\'{e}rale
de Lausanne (EPFL), CH-1015 Lausanne, Switzerland}

\author{B. Dalla Piazza}
\affiliation{Institute for Condensed Matter Physics, \'{E}cole Polytechnique Fed\'{e}rale
de Lausanne (EPFL), CH-1015 Lausanne, Switzerland}
 
 \author{S. Toth}
 \affiliation{Laboratory for Neutron Scattering and Imaging, Paul Scherrer Institut, CH-5232 Villigen PSI, Switzerland}
 
  \author{J. Pelliciari} 
 \affiliation{Swiss Light Source, Paul Scherrer Institut, CH-5232 Villigen PSI, Switzerland}
 
\author{R.~Fittipaldi}
\affiliation{CNR-SPIN, I-84084 Fisciano, Salerno, Italy}
\affiliation{Dipartimento di Fisica "E.R.~Caianiello", Universit\`{a} di Salerno, I-84084 Fisciano, Salerno, Italy}

\author{A.~Vecchione}
\affiliation{CNR-SPIN, I-84084 Fisciano, Salerno, Italy}
\affiliation{Dipartimento di Fisica "E.R.~Caianiello", Universit\`{a} di Salerno, I-84084 Fisciano, Salerno, Italy}

\author{N.~Kikugawa}
\affiliation{National Institute for Materials Science, 1-2-1 Sengen,  Tsukuba,  305-0047  Japan} 
\affiliation{National High Magnetic Field Laboratory, Tallahassee, Florida 32310, USA}

\author{J.~S.~Brooks}
\affiliation{National High Magnetic Field Laboratory, Tallahassee, Florida 32310, USA}

\author{H.~M.~R\o{}nnow}
\affiliation{Institute for Condensed Matter Physics, \'{E}cole Polytechnique Fed\'{e}rale
	de Lausanne (EPFL), CH-1015 Lausanne, Switzerland}
	\affiliation{Institute for Solid State Physics (ISSP), The University of Tokyo, Kashiwa, Chiba 277-8581, Japan}

\author{M. Grioni}
\affiliation{Institute for Condensed Matter Physics, \'{E}cole Polytechnique Fed\'{e}rale
	de Lausanne (EPFL), CH-1015 Lausanne, Switzerland}

 \author{Ch. R\"{u}egg}
 \affiliation{Laboratory for Neutron Scattering and Imaging, Paul Scherrer Institut, CH-5232 Villigen PSI, Switzerland}
\affiliation{Department of Quantum Matter Physics, University of Geneva, CH-12111 Geneva, Switzerland}

 \author{T.~Schmitt}
 \affiliation{Swiss Light Source, Paul Scherrer Institut, CH-5232 Villigen PSI, Switzerland}

 \author{J.~Chang}
 \affiliation{Institute for Condensed Matter Physics, \'{E}cole Polytechnique Fed\'{e}rale
 	de Lausanne (EPFL), CH-1015 Lausanne, Switzerland}
 \affiliation{Physik-Institut, Universit\"{a}t Z\"{u}rich, Winterthurerstrasse 190, CH-8057 Z\"{u}rich, Switzerland}
 
\begin{abstract}
	High-resolution resonant inelastic X-ray scattering (RIXS) at the oxygen $K$-edge
	has been used to study the orbital excitations of 
	\Ca\ and \Sr. In combination with  linear dichroism X-ray absorption spectroscopy, 
	the ruthenium $4d$-orbital occupation and excitations 
	were probed through their hybridization with the oxygen $p$-orbitals.
These results are described within a minimal model, taking into account 
 crystal field splitting and a spin-orbit coupling $\lambda_{so}=200$~meV.
The effects of spin-orbit interaction on the electronic structure and implications
for the Mott and superconducting ground  states of {(Ca,Sr)}$_2$RuO$_4$ 
are discussed.	 
\end{abstract}

\maketitle
\date{\today}

\maketitle
\section{Introduction}
The relativistic 
coupling between electronic spin and orbital momentum was 
long thought to have marginal influence on electrons in solids. 
Following the prediction and observation of topological 
surface states on Bi-based compounds~\cite{hasanRMP2010}, this paradigm has changed. 
Discovery of novel quantum phases realized through strong spin-orbit interaction 
is now a vivid field of research~\cite{krempaARCMP2014}. 
The demonstration of   
spin-orbit coupling driving a new 
type of Mott insulating state in layered 
iridates~\cite{bjkimPRL2008} is a good example of this.
It has been proposed that doping of this effective
$J_{1/2}$-Mott insulating state could lead to an exotic 
type of superconductivity~\cite{fawangPRL2011}, 
where Cooper pairs are composed of strongly spin-orbit coupled electrons.

In this context, it is interesting to study
other systems that display Mott physics and superconductivity in 
conjunction with strong spin-orbit interaction. %of which 
 The $4d$-transition metal oxide system Ca$_{2-x}$Sr$_x$RuO$_4$ represents such a case.
For $x=0$, the system  
is a Mott insulator, whose exact nature is not clarified~\cite{gorelovPRL2010,hottaPRL2001,jhjungPRL2003,jsleePRL2002,zegkinoglouPRL2005,KhaliullinPRL10,akbariPRB2014}. 
%It has, for example, 
%been proposed that the 2/3-filled $t_{2g}$-band becomes insulating through 
%an orbital-selective transition~\cite{gorelovPRL2010}. 
At the opposite stoichiometric end ($x=2$),
 the system has a superconducting ground 
state ($T_c=1.5$~K) originating from a correlated 
Fermi liquid~\cite{mackenzieRMP2003}. Although triplet $p$-wave superconductivity was
proposed early on,  the mechanism and symmetry class of the superconducting 
	order parameter is still debated~\cite{raghuPRB2010,riceJPCM1995,yyanaseJPSJPN2014,mazinPRL1997,mazinPRL1999}. 
	
	A fundamental question is 
how strongly spin-orbit interaction influences the electrons in 
these materials and whether it has an impact on the Mott insulating and superconducting ground states?
Current experimental evidence for a strong spin-orbit interaction 
stems from absorption spectroscopy~\cite{mizokawaPRL2001,malvestutoPRB2011,malvestutoPRB2013}, that has revealed a
considerable admixture of the ruthenium $t_{2g}$ orbitals.
More recently, spin-resolved photoemission spectroscopy has 
reported spin-polarized bands in \Sr~\cite{veenstraPRL2014,haverkortPRL2008}.
However, the most direct consequence of strong spin-orbit interaction -- 
the splitting of $t_{2g}$ states -- has not yet been  
probed directly by experiments. Orbital excitations transferred across this splitting 
are in fact not accessible to optical spectroscopies. 
Furthermore, the Ru $L$-edge~($\sim3$ keV)~\cite{zegkinoglouPRL2005} is 
currently inaccessable to high-resolution RIXS instrumentation (as it lies right between soft and hard X-ray optics).

To overcome these experimental challenges, we access here the Ru $4d$-orbital excitations through their hybridization with oxygen 
$p-$orbitals. 
Exploiting a combination of 
X-ray absorption (XAS) and oxygen $K$-edge resonant inelastic X-ray spectroscopy (RIXS), 
we provide direct evidence for a splitting of the ruthenium $t_{2g}$ states. % are provided. 
Our RIXS study of \Ca\ and \Sr~reveals excitations that allow an estimation of the spin-orbit 
coupling, in the same fashion as for the iridates~\cite{xliuPRL2012,salaPRB2014}. These results suggest 
a spin-orbit coupling $\lambda_{so}$ of $\sim 200$~meV -- only about two times weaker 
than in the iridates.  We conclude by discussing the Mott insulating 
and superconducting states in Ca$_{2-x}$Sr$_x$RuO$_4$.  \\

\section{Methods}
 High-quality single crystals of \Sr\ 
and \Ca\ were grown by the flux-feeding floating-zone technique~\cite{deguchiJPSJPN2002,snakatsujiJSSCHEM2001}.
The samples were aligned \textit{ex-situ} and cleaved \textit{in-situ} using the top-post method,
to access
momenta along the  Ru-O bond direction. Oxygen $K$-edge X-ray absorption spectroscopy (XAS) and resonant inelastic X-ray scattering 
(RIXS) experiments were carried out at the ADvanced RESonant Spectroscopy (ADRESS) beamline at 
the Swiss Light Source (SLS)~\cite{strocovJSYNRAD2010,ghiringhelliREVSCIINS2006}. Absorption spectra were measured in fluorescence-yield mode, using both horizontally 
and vertically polarized light. The RIXS spectrometer was set to have a fixed scattering angle of 130 degrees and an 
energy resolution of $29$ meV (HWHM) at the oxygen $K$-edge. All spectra were recorded at 
$T=20$ K. XAS matrix elements and RIXS momentum $Q=(h,k,l)$ were varied by changing the incident angle $\theta$  
 (see inset Fig.~2(a)).\\[0mm]

\begin{figure}
\begin{center}
\includegraphics[width=0.42\textwidth]{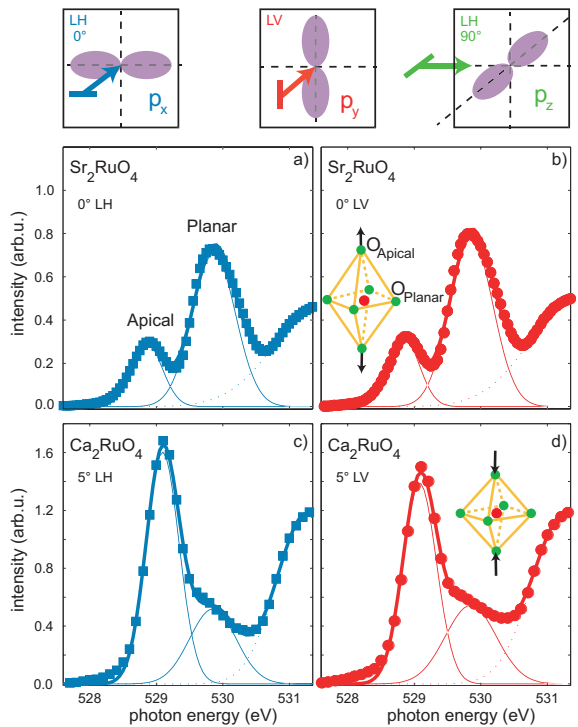}
\caption{(color online)  X-ray absorption spectra (XAS) of Sr$_2$RuO$_4$ (top) and \Ca (bottom) recorded using horizontal (left) 
and vertical (right) linearly polarized light near normal incidence ($\theta\sim 0^\circ$). A sloping 
background has been subtracted and solid lines are Gaussian fits. 
Top panels show schematically the oxygen $p_x$, $p_z$, and $p_y$ 
orbitals and how the cross section is optimized with 
different conditions of incident photon angle and polarization. Lower insets 
show the elongated and compressed octahedron. }\label{fig:fig1}
\end{center}
\end{figure}

\begin{figure}
\begin{center}
\includegraphics[width=0.4\textwidth]{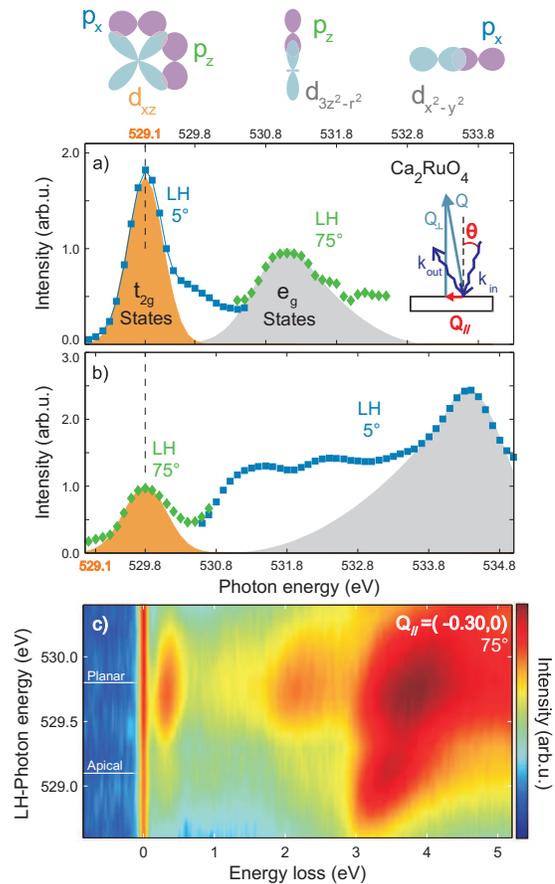}
\caption{(color online) \Ca. (a-b) X-ray absorption spectra recorded with linear horizontal
light and incident angle $\theta=5^\circ$ (blue squares) and $75^\circ$ (green diamonds).
(a) Absorption spectra mainly probing apical oxygen orbitals,
whereas XAS shown in  (b) are mostly sensitive to planar oxygen states.
Intensities in (a) and (b) are normalized to give approximative overlaps 
between the two curves.
Top panels schematically show the hybridization between Ru $d-$states 
and oxygen $p_{x/y}$ or $p_z$ orbitals (top). Inset in (a) illustrates the scattering geometry, the incident angle $\theta$ and momentum transfer $Q$. 
(c) Resonant inelastic X-ray spectra collected with momentum transfer and polarization 
as indicated and  displayed using a logarithmic color scale as a function 
of incident photon energy.  Notice that the photon energies in (b) are shifted relatively to (a), so that both apical and planar $t_{2g}$ resonances are aligned.   
 Furthermore, the elastic line was aligned to the XAS $t_{2g}$ resonances in (a) and (b), to allow a direct comparison between RIXS and XAS features.}
\label{fig:fig2}
\end{center}
\end{figure}

\section{Results}
In Fig. 1 and 2, X-ray absorption spectra recorded on \Ca\ and \Sr\
are shown for different light polarizations and incident angles $\theta$.
Good agreement with previous XAS work~\cite{malvestutoPRB2011,mizokawaPRL2001} is found whenever overlap in temperature, 
polarization, and incident angle is present.
As generally observed on cuprates~\cite{ctchenPRL1991}, iridates~\cite{salaPRB2014} and 
ruthenates~\cite{mizokawaPRL2001,malvestutoPRB2011}, the $t_{2g}$ and $e_g$
states can be probed through oxygen-hybridization on both the apical and 
planar oxygen sites, that have slightly different absorption resonance energies~\cite{ctchenPRL1991}.
 
 By varying light polarization and incident 
angle $\theta$, matrix elements favor different $p$-orbitals -- see top panels of Fig.~1.
 Linear vertical light is, independent of incidence angle, mostly sensitive to the oxygen 
$p_y$ orbitals. By contrast,
linear horizontal light predominantly probes $p_x$ orbitals for $\theta=0^\circ$ and $p_z$ for $\theta\sim 90^\circ$. % -- see top insets of Fig.~1. 
The degree of hybridization between 
the ruthenium (Ru$^{4+}$) $4d$-orbitals 
and the oxygen $p_x$,$p_y$, and $p_z$ orbitals also enters into the 
absorption cross section.
Therefore, varying incident angle $\theta$ and polarization on both planar and apical oxygen 
edges yields information about both ruthenium $e_g$- and $t_{2g}$-states.

The first two features in Fig. 1(a-d) -- appearing just below 530~eV -- are the oxygen-$K$ absorption resonances
due to Ru $t_{2g}$-hybridization with apical and planar oxygen, respectively~\cite{mizokawaPRL2001,malvestutoPRB2011}. 
Features at higher energies are attibuted to hybridization with $e_g$~($d_{3z^2-r^2}$ and $d_{x^2-y^2}$)~Ru orbitals~\cite{mizokawaPRL2001}.
The $d_{3z^2-r^2}$ states are best probed through $p_z$ hybridization 
on the apical oxygen site -- see Fig. 2(a). Comparing $t_{2g}~(d_{xy},d_{xz},d_{yx})$  and $d_{3z^2-r^2}$ apical absorption 
resonances  
suggests a splitting of 
approximately $2$~eV in \Ca. In a similar fashion, $d_{x^2-y^2}$
states are best probed through the planar oxygen sites. 
There, the $t_{2g}$ to $d_{x^2-y^2}$ splitting  (Fig. 2(b)) 
is $\sim3-4$~eV. Comparable energy scales were found in \Sr.
   
\begin{figure}
	\begin{center}
		\includegraphics[width=0.44\textwidth]{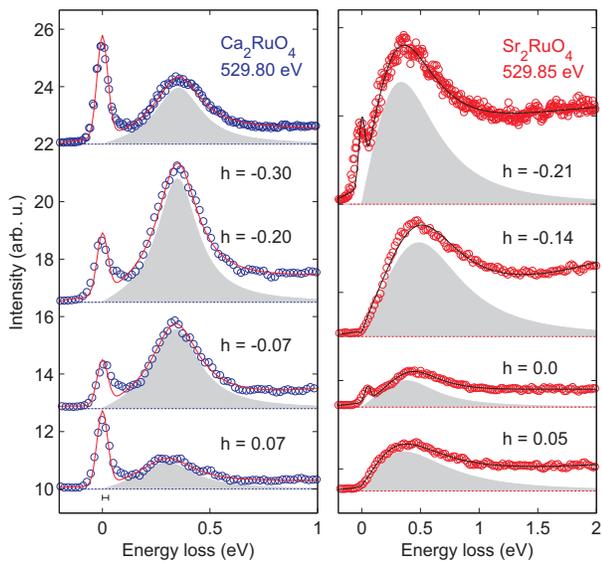}
		\caption{(color online) Resonant Inelastic X-ray Scattering (RIXS) spectra for different momentum transfers $Q_{||}=(h,0)$ as 
			indicated on \Ca\ (left) and \Sr\ (right),  
			recorded using linear-horizontal light tuned 
			to the planar oxygen $K$-edge. For visibility, all 
			spectra are given an individual vertical offset. Solid lines are 
			fits to a Gaussian (approximately resolution limited elastic line), an anti-symmetric Lorentzian (gray-shaded 
			component) and a quadratic background. 
		}\label{fig:fig3}
	\end{center}
\end{figure}

\begin{figure}
	\begin{center}
		\includegraphics[width=0.34\textwidth]{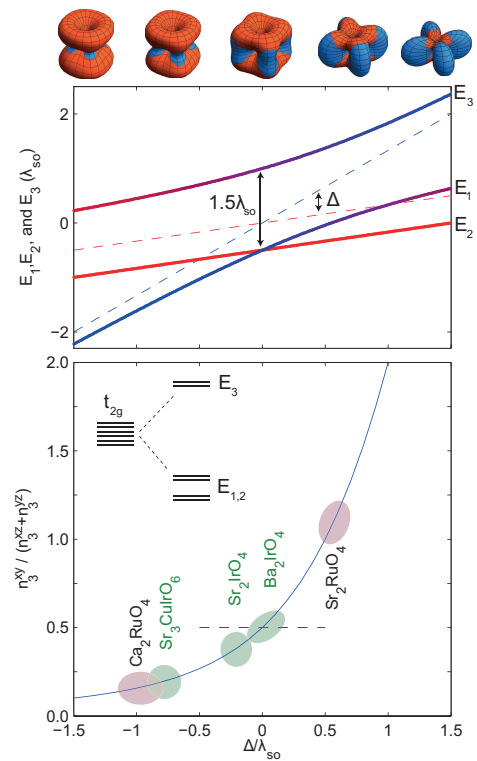}
		\caption{(color online) (a) Eigenenergies $E_2$,$E_1$, and $E_0$ of the Hamiltonian given in Eq.~1 versus 
			$\Delta/\lambda_{so}$, where $\Delta$ is the 
			crystal field splitting of the $t_g$ states and $\lambda_{so}$ is the 
			spin-orbit coupling strength. 
			Dashed (solid) lines are the solutions in absence (presence) of spin-orbit interaction. Color indicates 
			the orbital character with blue being $d_{xy}$ and red being $d_{yz}$ or $d_{xz}$. Top-panels display the orbital 
			topology of the $E_2$ Eigenstate. (b) Ratio $n^{xy}_3/(n^{xz}_3+n^{xz}_3)$ between orbital occupation of 
			$d_{xy}$ and $d_{xz}$ versus $\Delta/\lambda_{so}$. The solid line is the model expectation of 
			the Hamiltonian given in Eq.~\ref{eq:eq1}. Values of $\Delta/\lambda_{so}$ for the iridate
			materials stem 
			from Refs.~\onlinecite{salaPRB2014,bjkimPRL2008,jkimNATCOMM2014,xliuPRL2012}. The inset shows schematically how the $t_{2g}$
			states are split by spin-orbit interaction and crystal field.
		}\label{fig:fig4}
	\end{center}
\end{figure}

\section{RIXS} 
Next, we turn to the resonant inelastic X-ray spectra 
recorded on \Ca\ and \Sr. In Fig. 2(c), the incident-photon-energy dependence of 
the RIXS spectra across apical and planar oxygen $K$-resonances 
on \Ca\ is shown for linear horizontal light polarization at
incident angle $\theta=75^\circ$. 
Besides elastic scattering, three pronounced excitations are resolved 
at the planar oxygen edge. Those at $\sim 2$ and $\sim 4$~eV, correspond approximately
to the $t_{2g}$ to $d_{3z^2-r^2}$ and  $d_{x^2-y^2}$ splittings
and are hence assigned to be $dd$-excitations. 
In the following, focus is on the 
 low-energy excitations found at $0.3-0.5$~eV  
for both \Ca\ and \Sr -- see Fig.~3. 
These excitations are non-dispersive
and reside at energies well above optical phonon branches.
Furthermore, 
as \Ca\ is an insulator, a plasmon scenario is very unlikely.
These are also incompatible with a simple $t_{2g}$ crystal field 
splitting, which is expected to be much smaller than 300~meV.

\section{Interpretation} 
To gain further insight into the nature of this 
excitation, we start by discussing 
the $t_{2g}$ states. Linear dichroism effects on X-ray absorption spectra 
yield information about the orbital hole occupation $n^{xz}_3=n^{yz}_3$ and $n^{xy}_3$~\cite{mizokawaPRL2001,salaPRB2014}.
For example, on the planar oxygen site, $p_{x/y}-d_{xy}$  and $p_z-d_{xz/yz}$ hybridizations 
are dominating, whereas $p_{x/y}-d_{xz/yz}$ is leading  at the apical site.
Using  light polarization to emphasize the $p_x$ or $p_y$ channel, absorption 
is enhanced on the apical site if  hole orbital occupation $n^{yz}_3=n^{xz}_3$ is high.
Likewise, the planar absorption resonance will be enhanced for large $d_{xy}$-occupation.
As a result, apical and planar absorptions cannot both be strong at the same 
time. 

The proportion between planar and apical XAS peak amplitudes is an experimental measure of  
the ratio $R_3=n^{xy}_3/(n^{xz}_3+n^{yz}_3)$~\cite{mizokawaPRL2001,salaPRB2014}. 
Judging from peak amplitudes~\cite{mizokawaPRL2001}, $R_3\sim 1.23(2)$  and $\sim 0.17(2)$ respectively in \Sr\ and 
 \Ca\ -- see Fig.~4(b). There is, however, a caveat related to the tetragonal distortion of the apical oxygen 
(6$\%$ and $-2\%$ in \Sr\ and \Ca~respectively) leading to  
slightly under- and over-estimation of $n^{xz/yz}_3$~\cite{salaPRB2014}. 
Assuming (as done for iridates materials~\cite{salaPRB2014}) that the 
hybridization strength decays as $r^{-3.5}$~\cite{harrisonBOOK1980}, where $r$ is the 
Ru-O bond length, $n^{xz/yz}_3$ would be overestimated by $\sim 20\%$ in \Sr\ and 
underestimated by 5$\%$ in \Ca. Therefore,  $1< R_3< 1.25$ for \Sr\ 
and $0.15< R_3 < 0.2$ for \Ca\ (Fig. 4) -- the latter being consistent 
with the conclusion of early XAS work at 90 K using circular polarized light~\cite{mizokawaPRL2001}.

\section{Model} 
This mixing of 
\ketxy, \ketyz, and \ketxz orbitals, where $\pm$ refers to the electronic spin, 
can be explained by 
	a non-negligible spin-orbit interaction $\lambda_{so}$~\cite{haverkortPRL2008,veenstraPRL2014,gqliuPRB2011,gqliuPRB2013,mizokawaPRL2001}. 
Calculations including crystal field effects and spin-orbit interaction but neglecting 
the Hund's coupling~\cite{georgesARCMP2013}  
 have described very successfully  the band structure of \Sr\ and Sr$_2$RhO$_4$~\cite{haverkortPRL2008,veenstraPRL2014}.  
Following this spirit, 
the simplest Hamiltonian describing the $t_{2g}$  states reads:
\begin{equation}\label{eq:eq1}
 H=\lambda_{so} \mathbf{L}\cdot\mathbf{S}+\frac{\Delta}{3} \langle L_z\rangle^2
\end{equation}
where $\mathbf{S}$ and $\mathbf{L}$ are the spin and orbital momentum operators and $\lambda_{so}$
is the spin-orbit coupling constant~\cite{salaPRB2014,salaPRL2014,xliuPRL2012}.  
The intra-$t_{2g}$ crystal field splitting $\Delta$ %between $d_{xy}$- and $d_{yz},d_{xz}$- orbitals 
is defined so that $\Delta>0$ lifts $d_{xy}$ above $d_{xz}$ and $d_{yz}$.
Diagonalizing Eq.~1 in the (\ketxy, \ketyz, \ketxz)-subspace~\cite{salaPRL2014,xliuPRL2012} (neglecting $e_g$-states) yields 
 the eigenstates  \ketone$=$\ketxzmp$\pm i$\ketyzmp+$\sqrt{n^{xy}_1}$\ketxy, \kettwo$=$\ketxz$\mp i$\ketyz\ and  \ketthree$=$\ketxzmp$\pm i$\ketyzmp+$\sqrt{n^{xy}_3}$\ketxy\
with hole/electron occupancy:
\begin{equation}
 n^{xy}_3 =\frac{\left[2\delta-1+C\right]^2}{4} \quad \textrm{and} \quad  n^{xy}_1=n^{xy}_3-2C
\end{equation}
 where $\delta=\Delta/\lambda_{so}$ and $C=\sqrt{9+4\delta(\delta-1)}$ -- see Fig.~4. 
The Eigenenergies ($E_3,E_2$ and $E_1$) are split by: 
\begin{equation}
 E_3-E_1=\frac{\lambda_{so}C}{2}\quad\textrm{and}\quad E_3-E_2=\frac{\lambda_{so}}{4} (C+3+2\delta).
\end{equation}
Notice that in the limit $\delta\rightarrow0$ (the case of \{Ba,Sr\}$_2$IrO$_4$~\cite{salaPRB2014,bjkimPRL2008,jkimNATCOMM2014}),  
then $E_2=E_1$ are degenerate (see Fig.~4(a)) and $E_3-E_1=1.5\lambda_{so}$.
In the opposite limit $\lambda_{so}\rightarrow0$, $E_3-E_1 = \Delta$.
Within this simple model, our observables $R_3=n^{xy}_3/(n^{xz}_3+n^{yz}_3)$
and the RIXS excitation at $\sim350$~meV can be explained using the 
two adjustable parameters $\Delta$ and $\lambda_{so}$.

For example, 
for \Sr\ where $R_3\approx 1.2(2)$, 
we find $\Delta/\lambda_{so}\sim0.55(5)$ (see Fig.~\ref{fig:fig4}). 
This implies that $E_2-E_0=1.4\lambda_{so}$ and $E_2-E_1=2.1\lambda_{so}$.
Assuming that the peak feature at 
$\sim 350$~meV (Fig. 3(b)) results from the average of two broad excitations ($E_3-E_2$ and $E_3-E_1$) leads us to
$\lambda_{so}\sim 200$~meV and hence $\Delta\sim 100$~meV. 
This value of $\lambda_{so}$ is comparable to 
the theoretical expectation for Ru~\cite{veenstraPRL2014,hkontaniPRL2008} and 
what has been extracted from spin resolved ARPES~\cite{veenstraPRL2014}. 
Notice that the $E_2-E_1 \sim \Delta/2 \approx 50$~meV splitting -- possibly accessible 
through indirect RIXS processes -- is expected near the elastic line but 
not resolved in this experiment.

As $R_3\sim1/6$ for \Ca, it implies that $\delta\sim -1$ 
and hence $E_2-E_0=2.1\lambda_{so}$ and $E_2-E_1=1.3\lambda_{so}$. 
The RIXS spectra, shown in Fig.~3(a), exhibit a pronounced excitation at 
$\sim 340$~meV. If this is a result of an average of two excitations, 
once again $\lambda_{so}\sim 200$~meV is found. Thus, by including a spin-orbit coupling of 200 meV, 
 a consistent description of the orbital hole occupation extracted from XAS and 
 the excitations of the RIXS spectra on both \Sr\ and \Ca\ is obtained.

\section{Discussion} 
Implications of spin-orbit coupling $\lambda$ in $4d$-transition oxide materials
have already been evaluated in a number of papers~\cite{hkontaniPRL2008,KhaliullinPRL10,akbariPRB2014,haverkortPRL2008,yyanaseJPSJPN2014,gqliuPRB2011,gqliuPRB2013,iwasawaPRL2010}. 
The magnon bandwidth in \Ca\ is, for example, predicted~\cite{KhaliullinPRL10,akbariPRB2014} to 
be controlled  by $\sim 3\lambda /4$. Neutron experiments 
should be performed to test this prediction. Magnetic moments are also 
influenced but not uniquely defined by $\Delta/\lambda$~\cite{akbariPRB2014}.
%For example, it has been shown that the
%spin-orbit coupling improves significantly the agreement between calculated 
%and observed band structure~\cite{haverkortPRL2008,veenstraPRL2014}.
%We start by discussing the influence that spin-orbit coupling may have 
%on the Mott-insulating state of \Ca.
As the experiments on \Ca\ suggest that \ketthree\ is dominated by $d_{xz}/d_{yz}$ 
orbitals, it is possible to approximate  \ketthree$\approx$\ketxzmp$\pm i$\ketyzmp. Then, 
both \kettwo\ and \ketthree\ are more elegantly expressed in spherical harmonic 
notation:~\kettwo=$|\ell_z=\pm1 , s_z=\mp1/2>=\chi^{\pm 1/2}$ and \ketthree $= |\ell_z=\pm1 , s_z=\pm1/2>=\chi^{\pm 3/2}$.
In this simplistic limit,
the role of spin-orbit interaction is  to split the four-fold
degeneracy of \ketxz\ and \ketyz\ into two-fold
degenerated $\chi^{\pm 3/2}$ and $\chi^{\pm 1/2}$ states -- see Fig.~4.
It has been argued that even modest Coulomb interaction 
$U$ is sufficient to split these $\chi^{\pm 3/2}$ and $\chi^{\pm 1/2}$ states
and hence drive the Mott insulating transition~\cite{gqliuPRB2011,gqliuPRB2013}.
Therefore, as in layered iridates, a combination of spin-orbit interaction 
and electron correlations may be sufficient to drive the 
Mott insulating ground state.

Another interesting question is how spin-orbit interaction impacts 
the superconducting ground state in \Sr. It has been suggested 
theoretically that ferromagnetic interactions  would result 
in a chiral $p$-wave superconducting state~\cite{riceJPCM1995} driven by the $d_{xy}$-dominated $\gamma$-band. By contrast, if superconductivity is driven 
by the $d_{xz}/d_{yz}$-dominated $\alpha$- and $\beta$-bands~\cite{raghuPRB2010}, then spin-orbit coupling lifts the 
ground state degeneracy 
in favor of a helical $p$-wave symmetry~\cite{yyanaseJPSJPN2014}. These considerations were, however, based 
on the assumption that spin-orbit interaction is weak compared to 
the Fermi energy $E_F$~\cite{yyanaseJPSJPN2014}.  It is hence useful to compare 
the energy scales of superconductivity, spin-orbit coupling and the Fermi 
energy $E_F$. As $T_c=1.5~$K, the superconducting gap amplitude 
is expected in the $\sim1$~meV range~\cite{firmoPRB2013}.
The Fermi energy $E_F=[\hbar/(4\pi k_B)] (A_k / m^*)$~\cite{jchangPRL2010} can be 
estimated from the Fermi surface area $A_k$ and the quasiparticle mass $m^*$.
For the $\gamma$-band, quantum oscillation experiments~\cite{mackenzieRMP2003} yield oscillation frequency 
 $\hbar A_k / 2e \pi = 18$~kT and $m^*=16 m_e$, where $m_e$
is the free electron mass. These values imply that $E_F\sim 150$~meV, and as expected 
$k_B T_c / E_F \ll 1$. Similar values of $E_F$ are found for the $\alpha$- and 
$\beta$- bands. Strong electron correlations therefore drive 
even the $\gamma$-electrons into the regime $\lambda_{so}\sim E_F$,
where spin $S_z$ and orbital $L_z$ are no longer good quantum numbers.
Cooper pairs in \Sr\  therefore have to be 
composed of electronic pseudo-spins.
If realized, the same would likely be true for superconductivity 
in layered iridates.

\section{Conclusions and outlook}
In summary, we have performed a combined light absorption and oxygen $K$-edge resonant inelastic X-ray 
spectroscopy study of the ruthenates (Ca,Sr)$_2$RuO$_4$. 
Special attention was given to the Ru $t_{2g}$ states, probed through their hybridization with oxygen $p$-orbitals.
Both the oxygen $K$-edge RIXS and absorption spectra find a
consistent description within a simple model that includes crystal field splitting and 
 spin-orbit coupling $\lambda_{so}\approx 200$~meV. In
this picture, the main new observation --  RIXS excitations 
at $\sim350$~meV -- is interpreted as holes moving across spin-orbit 
split $t_{2g}$ states.

\textit{Acknowledgments: } 
We thank M. Sigrist and V.~M.~Katukuri for enlightening discussions. This research is funded by the Swiss National Science Foundation
and its Sinergia network Mott Physics Beyond
the Heisenberg model (MPBH).
 Experiments have been performed at the ADRESS beamline of the Swiss
Light Source at Paul Scherrer Institut. JP and TS acknowledge financial support through the Dysenos AG by Kabelwerke Brugg AG Holding, Fachhochschule Nordwestschweiz and the Paul Scherrer Institut.

\bibliographystyle{apsrev4-1}

%\section{}

\end{document}